\documentstyle[aps,multicol,epsfig]{revtex}

\begin{document}

\title{The metallic state in disordered quasi-one-dimensional conductors}

\author{H.C.F. Martens, J.A. Reedijk, and H.B. Brom}

\address{Kamerlingh Onnes Laboratory, Leiden University, P.O. Box\\
9504, 2300 RA Leiden, The Netherlands}

\author{D.M. de Leeuw}

\address{Philips Research Laboratories, Professor Holstlaan 4, 5656 AA\\
Eindhoven, The Netherlands}

\author{R. Menon}

\address{Department of Physics, Indian Institute of Science, Bangalore,\\
560012 India}

\date{\today}

\maketitle

\begin{abstract}
The unusual metallic state in conjugated polymers and single-walled
carbon nanotubes is studied by dielectric spectroscopy (8--600~GHz). 
We have found an intriguing correlation between scattering time and 
plasma frequency. This relation excludes percolation models of the
metallic state. Instead, the carrier dynamics can be understood in 
terms of the low density of delocalized states around the Fermi level, 
which arises from the competion between disorder-induced localization
and interchain-interactions-induced delocalization.\\

\end{abstract}


\begin{multicols}{2}
\settowidth{\columnwidth}{aaaaaaaaaaaaaaaaaaaaaaaaaaaaaaaaaaaaaaaaaaaaaaaaa}

The finite conductivity of a metal at zero Kelvin is a consequence of the
lattice-periodicity and the finite density of states at the Fermi level ($%
E_{F}$). By breaking translational symmetry, disorder localizes
charge-carriers. Upon growing disorder, eventually a metal-insulator
transition (MIT) occurs\cite{Anderson58}, the effect being the stronger the 
lower the dimensionality. In one dimension (1D) any disorder localizes 
the electronic
states. The MIT in quasi-1D conducting polymers and single-walled carbon
nanotubes is also disorder-driven, but its exact nature is under severe
debate. Many authors claim the presence of a ``heterogeneous'' state in
which the relevant disorder length-scale is large compared to the electronic
correlation-length. In this case, the MIT corresponds to a percolation
transition of metallic islands embedded in an amorphous matrix.
\cite{Kohlman97,Joo98,Kaiser98}
Other studies suggest that the MIT is of the Anderson type\cite{Anderson58}
with disorder occurring on length-scales equal or less than the electronic
correlation-length.\cite{Yoon94,Lee95} Then, extended and localized states are
separated in energy by the mobility edge ($E_{c}$), and the MIT occurs when $%
E_{F}$ crosses $E_{c}$.

\begin{figure}[tbf]
\hbox{\psfig{figure=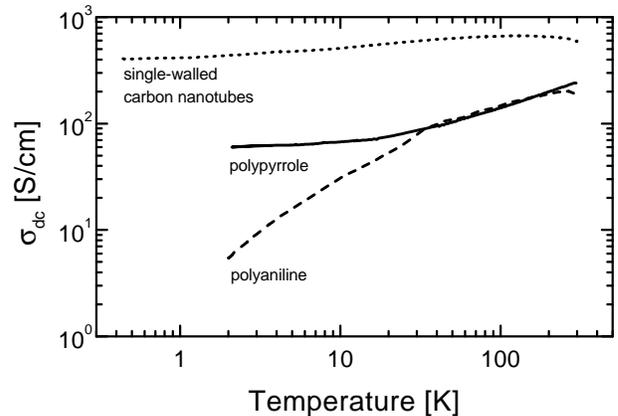,width=8cm}}
\caption{Temperature dependent dc conductivity of three disordered quasi-1D
systems. All samples are close to the MIT. Below 10~K the dc conductivities
of polypyrrole and single-walled carbon nanotubes are almost
temperature-independent implying a non-zero conductivity at zero Kelvin as
expected for a metal. The dc conductivity of polyaniline vanishes when
cooling, indicating that this sample is on the insulating side of the MIT.}
\label{dc}
\end{figure}

We have studied charge transport in polyaniline, polypyrrole, and
single-walled carbon nanotubes. Preparation details are given elsewhere.
\cite{Yoon94,Martens99,Hilt00} 
The temperature-dependent dc conductivity is shown in Fig.~\ref{dc}. 
Both single-walled carbon nanotubes and polypyrrole have a finite dc
conductivity down to the lowest temperatures, indicating a metallic state.
The dc conductivity of polyaniline vanishes when cooling, characteristic of
an insulating phase. Clearly, all samples are on the boundary of the MIT.
Unfortunately, these data alone are not sufficient to discriminate between
the above mentioned models.

Optical experiments have been widely used to study disordered quasi-1D
conductors.\cite{Kohlman97,Lee95,Martens99,Hilt00,Ugawa99,Chapman99} 
For conventional metals, the frequency ($\omega=2\pi f$) dependence of the 
complex conductivity, $\sigma^*=\sigma + i\omega\varepsilon_0\varepsilon$ 
($\varepsilon_0$ vacuum permittivity), is well explained in terms of the 
Drude free-electron model: 
\begin{equation}
\sigma^*(\omega) = {\frac{{\varepsilon_0\omega_p^2\tau}}{{1+i\omega\tau}}}
\label{Drude}
\end{equation}
with $\tau$ the scattering time and 
\begin{equation}
\omega_p=\sqrt{ne^2/(\varepsilon_0m^*)}\label{wp}
\end{equation}
the unscreened plasma frequency; $n$ free-carrier density, $e$ electronic charge, 
and $m^*$ effective mass. For $\omega\tau<1$, the dielectric constant ($%
\varepsilon$) is negative and the conductivity ($\sigma$)
frequency-independent. For $\omega\tau>1$, $\sigma$ drops to zero, while $%
\varepsilon$ increases and eventually becomes positive above $\omega_p$. For
normal metals $\omega_p\sim$~1--10~eV, and $\tau\sim10^{-14}{\rm \ s}$.
\cite{Ordal83} Free-carrier absorption ($\varepsilon<0$ in the microwave and
far-infrared regime) has been observed in polypyrrole\cite{Kohlman97}, 
polyaniline\cite{Kohlman97,Martens99}, and single-walled carbon nanotubes.\cite{Hilt00}
Here $\sigma^*(\omega)$ is examined in the range 8--600~GHz 
(0.27--20~cm$^{-1}$, 0.033--2.5~meV) by means of complex-dielectric 
spectroscopy\cite{Hilt00,Martens99,Reedijk00}. This technique covers
both the microwave and far-infrared regime, and does not rely on
Kramers-Kronig analyses.

\begin{figure}[tbf]
\hbox{\psfig{figure=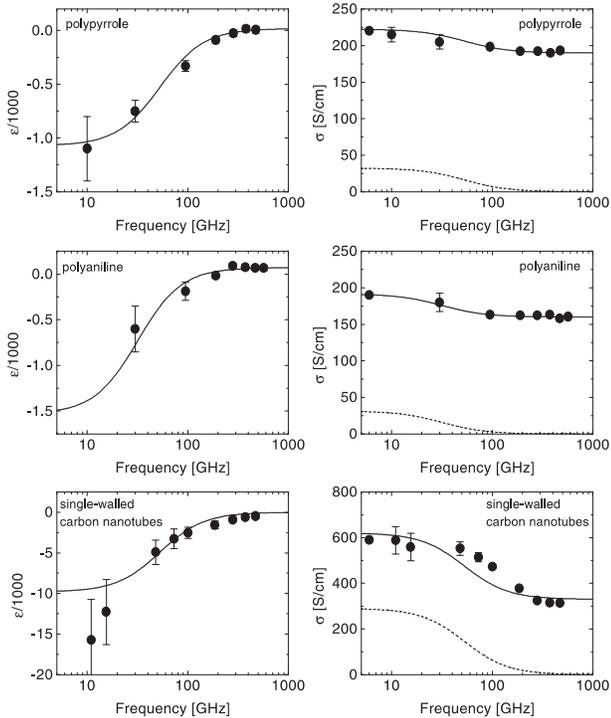,width=8cm}}
\caption{Room temperature dielectric function (left) and conductivity
(right) as a function of frequency for polypyrrole, polyaniline, and
single-walled carbon nanotubes respectively. In view of the logarithmic
scale, the value of $\sigma_{{\rm dc}}$ is plotted at $f=6$~GHz. The drawn
lines are fits to the data using the Drude equation with an extra (frequency
independent) background conductivity and dielectric constant. The dielectric
data are dominated by the free-carrier response. However, the free-carrier
contribution to the conductivity (dashed lines) is less than 50\%.}
\label{fits}
\end{figure}

For all samples, at low frequency $\varepsilon\ll 0$ as expected for a
metal, see Fig.~\ref{fits}. However, contrary to Eq.~(\ref{Drude}), $\sigma$
does not drop to zero but reaches a ``plateau'' at high frequency,
indicating an additional absorption mechanism. By incorporating a
frequency-independent background conductivity $\sigma_b$ and dielectric
constant $\varepsilon_b$ in Eq.~(\ref{Drude}) the data can be excellently
reproduced, see the solid lines in Fig~\ref{fits}. From the fits we find for
respectively polypyrrole, polyaniline, and single-walled carbon nanotubes: $%
{\rm \omega_p\ (meV)=7.3\pm0.5}$, $6\pm1$ and $22\pm7$; ${\rm \tau\
(ps)=3\pm 0.5}$, $5\pm1$ and $3\pm1.5$; ${\rm \varepsilon_b=18\pm 1}$, $%
70\pm10$ and $0\pm500$; ${\rm \sigma_b\ (S/cm)=190\pm10}$, $160\pm10$ and $%
370\pm50$. The free-carrier response of polyaniline shows that, for a sample
just on the insulating side of the MIT, extended states become thermally
occupied at finite temperature\cite{note}.

\begin{figure}[tbf]
\hbox{\psfig{figure=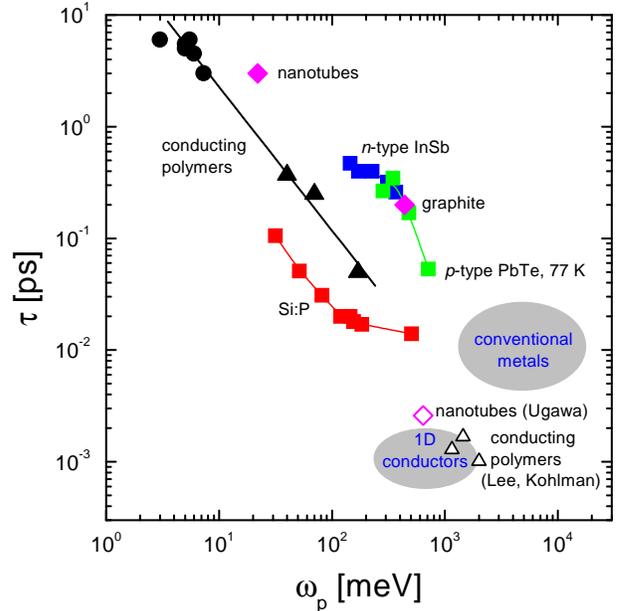,width=8cm}}
\caption{Room temperature values of the Drude parameters $\tau$ and $%
\omega_p $ describing the metallic state in doped conjugated polymers and
single-walled carbon nanotubes. For comparison the typical values of
conventional metals, crystalline 1D conductors, graphite, and several doped
semiconductors are indicated. In our view, the closed symbols correspond to
the free-carrier response of 3D extended states, while the open symbols can
be attributed to on-chain (1D) motion of charge-carriers. The empirical
correlation $\tau\propto\omega_p^{-1.3}$ for the conducting polymers, seems
to extrapolate to parameter ranges of conventional conductors. }
\label{trend}
\end{figure}

Fig.~\ref{trend} displays the room temperature Drude parameters of
disordered quasi-1D conductors, conventional metals\cite{Ordal83}, crystalline 1D
conductors\cite{1Da,1Db}, graphite\cite{Philipp77}, and doped semiconductors.
\cite{Spitzer57,Dixon65,Gaymann95}
The conducting polymers are given in black (dots this work, triangles
Ref.~2), and reveal a remarkable empirical correlation $\tau\propto%
\omega_p^{-1.3}$. Comparable trends are observed for the
doped-semiconductors. The open symbols correspond to the ``second'' plasma
frequency observed in conducting polymers\cite{Kohlman97,Lee95}, and 
single-walled carbon nanotubes\cite{Ugawa99}. The conducting polymers and 
single-walled carbon nanotubes
are very unlike conventional metals and crystalline 1D conductors. The
scattering times are surprisingly long, and counter-intuitively even
increase for the more disordered and ``less metallic'' samples.

\vspace{5mm}

The unusual carrier dynamics in conducting polymers have been argued to
indicate a quasi-1D percolating metallic state.\cite{Kohlman97} In this 
model, the long $\tau$ is explained in terms of one-dimensionality 
and the low $\omega_p$ as stemming from the small density of crystalline 
metallic regions. It was argued that, upon decreasing disorder, the fraction 
of metallic regions increases and this should increase $\omega_p$\cite{Kohlman97}.
However, the intrinsic conductive properties of metallic islands are not expected to 
depend on the 
concentration of such islands. For instance, in a bulk metal 
$\varepsilon$ is zero at the plasma frequency. Based on effective medium
calculations, Stroud\cite{Stroud79,BergmanStroud92} showed that the zero 
in $\varepsilon$ at $\omega=\omega_p$ persists in a metal-insulator-composite
for metal fractions above the percolation threshold, hence the plasma frequencies in 
the bulk and composite are the same. This is a natural consequence of the 
fact that $\omega_p$ only depends on the carrier density inside the 
percolating metallic path, and not on the free-carrier density in the 
total volume of the composite material. Indeed, recent experiments
on thin quench-condensed Pb films demonstrated that
$\omega_p$ in heterogeneous Pb films is independent of the fraction 
of Pb and almost equal to the plasma frequency of bulk lead\cite{Henning99}. 
In contrast, the plasma frequency in conducting polymers shows an increase of 
almost two orders of magnitude, see Fig.~\ref{trend}, at variance
with the behavior of a percolating metallic network.
Also, in terms of the heterogeneous model, Fig.~\ref{trend} would
imply that an increase of the fraction of metallic islands enhances the carrier 
scattering, which seems unlikely.
Based on the above theoretical and experimental findings, we conclude that 
the unusual carrier-dynamics in quasi-1D conductors can not be explained 
in terms of a percolating fraction of highly conducting regions, but rather 
reflects the intrinsic transport properties of the weakly metallic state in 
these materials.

The similarity between conducting polymers and doped semiconductors, 
Fig.~\ref{trend}, provides a strong clue that the free-carrier dynamics in 
these materials is governed by a common mechanism. In doped semiconductors 
all carriers are delocalized, $n=n_{{\rm doped}}\sim10^{23}$--%
$10^{25}{\rm \ m^{-3}}$ \cite{Spitzer57,Dixon65,Gaymann95}. In this 
three-dimensional, marginally metallic state the low 
$\omega_p$'s arise from the low band-filling. In conducting polymers 
$n_{{\rm doped}}\sim10^{27}{\rm \ m^{-3}}$, the high band-filling should 
give $\omega_p\sim1{\rm \ eV}$, but the observed plasma frequencies are 
orders of magnitude lower. Apparently, only a fraction of the carriers are 
delocalized ($n\ll n_{{\rm doped}}$).

\begin{figure}[tbf]
\hbox{\psfig{figure=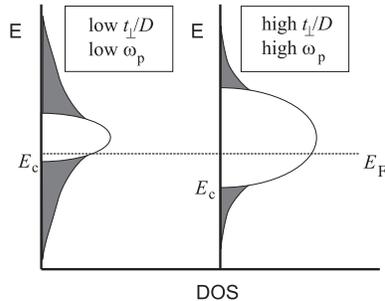,width=5cm}}
\caption{ Schematic drawing of the density of states (DOS) in disordered
quasi-1D conductors as a function of $t_\perp$ (interchain overlap) and $D$
(disorder bandwidth). Localized states (white) are separated in energy 
by the mobility 
edge $E_c$ from the extended states (gray). At constant doping level, when
increasing the ratio $t_\perp/D$ more extended states are formed, which
gives a larger free-carrier density $n$ and consequently a larger $\omega_p$. 
The metallic state only occurs when $E_c<E_F$ which requires a high doping
level together with large enough ratio $t_\perp/D$.}
\label{model}
\end{figure}

The low dimensionality of these systems enhances disorder-induced
localization. To obtain three dimensional (3D) extended states, inter-chain
charge transfer is a prerequisite. The amount of 3D extended states will be
governed by the competition between inter-chain overlap $t_{\perp }$ and the
strength of the disorder potential $D$. This is schematically depicted in 
Fig.~\ref{model}, less disorder corresponds to increasing $t_\perp/D$
and this increases the amount of delocalized states in the band.
In this model, 
localized and delocalized carriers are not spatially separated, but 
are separated in energy by the mobility edge.
As shown in Fig.~\ref{model}, even
if $n_{{\rm doped}}$ remains constant, both $n$ and $\omega _{p}$ will
increase when $t_{\perp }/D$ increases. In order to achieve
a metallic state, the doping level must be high enough to have $E_F$ in
the region of delocalized states, which explains why the metallic
state only occurs in highly doped polymers. Thus, the low $\omega _{p}$'s 
in disordered quasi-1D conductors reflect that, due to small inter-chain
overlap and strong disorder, the density of delocalized states is low, 
$n\ll n_{\rm doped}$.

To quantify the above discussion, we use Eq.~(\ref{wp}) and take as a 
typical density of states $g\sim$~1 state/(eV ring) \cite{Kohlman97} and 
$\omega_p\sim 30{\rm \ meV}$. As an estimate of $n$ we consider a
weakly metallic sample for which $E_F-E_c<k_BT$, so $n\approx k_BTg$, 
giving $n\sim10^{25}{\rm \ m^{-3}}$ and $m^*\sim15m_e$ ($m_e$ electronic mass).
Alternatively, when using a free-electron approximation, we find 
$n\sim10^{24}-10^{25}{\rm \ m^{-3}}$ and $m^*\sim10m_e$. As expected, 
$n\ll n_{{\rm doped}}$. The high 
$m^*$ agrees with a low interchain overlap, which gives narrow electronic
bands and hence heavy masses. From the derived $m^*$ and $n$ we estimate 
$E_F-E_c$ to be only a few meV. Indeed, $k_BT\gg E_F-E_c$, and the
delocalized carriers behave classically. From the average velocity 
$\overline{v}=\sqrt{8k_BT/\pi m^*}\approx 3\times10^4{\rm \ m/s}$, and
$\tau\sim5\rm\ ps$ for $\omega_p\sim30\rm\ meV$, we find a
mean free path $\ell\sim10\rm\ nm$. Since the electronic de Broglie wavelength 
is comparable or less than the mean free path, the Drude analysis seems
justified.

In both single-walled carbon nanotubes and conducting polymers a second
plasma frequency has been reported around 1~eV, with $\tau\sim10^{-15}{\rm \
s}$.\cite{Kohlman97,Lee95,Ugawa99} These values match the on-chain 
parameters in crystalline 1D-conductors, and could reflect the motion of 
carriers which are not 3D-delocalized but confined to 1D chains. Lee 
{\it et al.} calculated that these carriers contribute 50--70\% to the total 
dc conductivity\cite{Lee95}, in agreement with $\sigma_b$ in the fits of the 
GHz response (Fig.~\ref{fits}).

The empirical interrelation between $\tau$ and $\omega_p$ seems generic for
marginally metallic systems with low free-carrier density. The decrease of 
$\tau$ upon doping semiconductors has been suggested to result from impurity
scattering.\cite{Gaymann95} However, for conducting polymers impurity or grain
boundary scattering is not dominant at room temperature, since this would
lower $\tau$ in the more disordered systems (lower $\omega_p$). In a
non-degenerate electron gas, due to screening, the electron-electron (e-e)
scattering cross section $\Sigma\propto n^{-1}$, and $\tau_{\rm e-e}\propto1/(%
\overline{v}n\Sigma)$ is independent of $n$. Apparently, at room temperature
electron-phonon (e-ph) scattering dominates. Since the e-ph scattering rate
is proportional to the density of states, an increase of $\omega_p$,
reflecting the growing density of delocalized states, naturally leads to
shorter $\tau$'s. Indeed, the empirical correlation extrapolates to
parameter ranges of conventional metals at room temperature, where e-ph
scattering is known to govern the carrier dynamics.

Finally, we address the role of heterogeneity
of the structure of disordered quasi-1D conductors, which lies at the heart
of the controversy between ``homogeneous'' and ``heterogeneous'' disorder 
models. It is well established 
that the metallic state in conducting polymers requires careful preparation
in order to minimize structural disorder. Diffraction experiments 
show that in the best metallic polymers the crystalline coherence length $\xi$ 
is at most several nanometers\cite{Kohlman97,Joo98,Nogami94}. Metallic 
islands could be formed when the crystalline coherence length 
is larger than the extent of electronic wavefunctions. However, from
the present work we find that $\xi<\ell$, and this implies that the
delocalized charge-carriers experience an avarage of the structural disorder. 
This corroborates our arguments that the carrier dynamics in quasi-1D
conductors is not in agreement with percolation of metallic islands.
In the model proposed here the role of structural (dis)order can be naturally 
explained. An increasing crystalline coherence length reflects that the
polymer chains are mutually better ordered. This will favor 
interchain interactions and, at the same time, reduce the amount of disorder
experienced by the charge carriers. Both effects enhance the metallic state. 

The unusual carrier dynamics in conjugated polymers and single-walled carbon 
nanotubes has been consistently explained in terms of a three-dimensional 
marginally metallic state. The empirical correlation between $\tau$ and $\omega_p$, 
which is observed in conventional, doped semiconductors and disordered quasi-1D
conductors, 
is a consequence of the phonon-scattering mechanism and low density of 
extended states around $E_F$ in these materials. In quasi-1D conductors the 
low free-carrier density can not be explained in terms of heterogeneity, but 
is governed by the competition between inter-chain charge 
transfer and disorder-induced localization onto 1D chains. 
Extrapolating the empirical correlation suggests the 
possibility of further improvement of the conductive properties of these
materials, though beyond those of conventional metals seems doubtful.

Discussions with L.J. de Jongh are gratefully acknowledged. This work is
part of the research programme of FOM.

\end{multicols}

\end{document}